\documentstyle[ltwol,epsf]{article}

\begin{document}

\title{TOWARD THE CHIRAL LIMIT OF QCD: \\
Quenched and Dynamical Domain Wall Fermions}

\author{PING CHEN, NORMAN CHRIST, GEORGE FLEMING, ADRIAN KAEHLER,
CATALIN MALUREANU, ROBERT MAWHINNEY, GABRIELE SIEGERT, CHENGZHONG
SUI, YURI ZHESTKOV}

\address{Department of Physics, Columbia University, New York, NY
10027}   

\author{PAVLOS VRANAS}

\address{Physics Dept., University of Illinois, Urbana, IL
61801}

\twocolumn[\maketitle\abstracts{A serious difficulty in
conventional lattice field theory  calculations is the coupling
between the chiral and continuum limits. With both staggered and
Wilson fermions, the chiral limit cannot be realized without
first taking the limit of vanishing lattice spacing. In this
talk, we report on extensive studies of the domain wall
formulation of lattice fermions, which avoids this difficulty at
the expense of requiring that fermion propagators be computed in
five dimensions.   A variety of results will be described for
quenched and dynamical simulations at both zero and finite
temperature.  Conclusions about the benefits of this new method
and some new physical results will be presented.  These results
were obtained on the {\it QCDSP} machine recently put into
operation at Columbia and the RIKEN Brookhaven Research Center.}]

\section{Introduction}

Important theoretical and algorithmic advances have opened a new
approach to the numerical study of chiral symmetry in QCD.  The
two widely studied lattice fermion formalisms, staggered and
Wilson fermions, both present serious difficulties to the
numerical simulation of chiral symmetry in lattice QCD.  Not only
do both of these lattice descriptions explicitly break most or
all of the chiral symmetries present in massless QCD, but they
also obscure the underlying relationship between topology in the 
gauge sector and zero modes for the fermions.  As a result,
lattice calculations will typically fail to show the full
consequences of the Goldstone theorem, current algebra, or the 't
Hooft solution to the $U_A(1)$ problem, without explicit
extrapolation to the continuum limit.  Thus, the physics of
chiral symmetry discovered in both the 60's and the 70's is
corrupted by the usual lattice discretization of QCD.

However, by considering Wilson fermions formulated in five
dimensions with a large negative mass, $m_{\rm W}=-m_0$, 
Kaplan \cite{KAPLAN} showed that it is possible to avoid the
fermion doubling problem while still achieving complete chiral
symmetry in the much simpler limit of large lattice extent in the
new fifth dimension.  A further important advance was achieved by
Narayanan and Neuberger \cite{NN} who generalized Kaplan's
approach and recognized that this limit of large fifth dimension
not only realized the desired continuum chiral symmetry but also
created a structure of exact fermion zero modes with a close
relation to the continuum Atiyah-Singer theorem and the physics
of the axial anomaly.  An efficient and well-elaborated version
of this domain wall method was developed and analyzed by Shamir
and by Shamir and Furman,\cite{SHAMIR1,SHAMIR2} providing a
practical and attractive method for large-scale numerical 
calculation.\cite{NEUBERGER}  There has now been considerable
exploratory numerical work suggesting that this method lives up
to its promise for both the Schwinger model \cite{VRANAS} and
quenched QCD \cite{BLUM-SONI} and that fermion zero-modes with
the desired properties are 
realized.\cite{NARAYANAN-VRANAS,EDWARDS}

We will describe these issues in somewhat more detail in Section
\ref{sec:DWF} below and then describe a series of calculations
\cite{RIKEN-BNL,ZMODES,LAT98-QUENCHED,LAT98-PBP,LAT98-THERMO,LAT98-TOPOLOGY} 
underway on the recently completed
computers at Columbia (8,196-nodes 0.4Tflops) and the RIKEN
Brookhaven Research Center (12,288-nodes, 
0.6Tflops).\cite{LAT98-QCDSP}  These calculations are designed to
i)  establish the extent to which zero-mode effects can be seen
in practical simulations with fifth-dimension extent $L_s \approx
10$ (Section~\ref{sec:zero-modes}), ii) study the chiral symmetry
of quenched QCD with careful control over the effects of $L_s$,
examining both the hadron spectrum and the chiral condensate
(Section~\ref{sec:zero-temp}) and iii) explore QCD thermodynamics
using domain wall fermions both as a probe of the properties of
the pure gauge theory and as a method to realize complete flavor
symmetry in a lattice study of the full QCD phase transition. 
This final calculation represents the best-controlled approach
yet available to examine the role of the axial anomaly in the
full QCD, chiral phase transition.

\section{Domain Wall Fermions}
\label{sec:DWF}

\subsection{Formulation}

Our formulation of domain wall fermions follows closely that
proposed by Shamir.\cite{SHAMIR1,SHAMIR2}  The starting point is
the standard lattice treatment of the gauge variables as 
$3 \times 3$ special unitary matrices $U_\mu(n)$, associated with
each link in a four-dimensional space-time lattice.  Here $n$
locates the site from which the corresponding link extends in the
positive $\mu^{th}$ direction.  These link variables enter the
Wilson gauge action in the usual way.   The new features of the
domain wall treatment appear in the fermion action.  Here the
fermion field, $\psi(n,s)$ is 4-spinor and 3-component color
vector as usual but now depends on both the normal lattice
coordinate $n$ and a new fifth coordinate $s$, lying in the range
$0 \le s \le L_s-1$.  The action takes the form:
\begin{eqnarray}
\nonumber
{\cal A}_{\rm DWF}&=&
\sum_{n,s;n^\prime,s^\prime}
\bar\psi(n,s)\{(D_{\rm W})_{n,n^\prime}\delta_{s,s^\prime} +m_0\\
&&\quad+(D_5)_{s,s^\prime}\delta_{n,n^\prime}\} \psi(n^\prime,
s^\prime)
\end{eqnarray}
Here $D_{\rm W}$ is the usual Wilson Dirac operator:
\begin{eqnarray}
(D_{\rm W})_{n,n^\prime} &=&{1 \over 2}\sum_\mu\bigl\{
(1+\gamma^\mu)U(n)_\mu\delta_{n,n^\prime-\hat\mu} \\
&+&(1-\gamma^\mu)U(n^\prime)_\mu^\dagger
                      \delta_{n,n^\prime+\hat\mu} -
2\delta_{n,n^\prime}\bigr\},
\nonumber
\end{eqnarray}
$m_0$ the Wilson mass (but with an unconventional negative sign),
and $D_5$ an additional fifth-dimension piece, diagonal in color:
\begin{eqnarray}
(D_5)_{s,s^\prime} &=& {1 \over 2}\bigl\{
(1+\gamma^5)\delta_{s,s^\prime-1} +
(1-\gamma^5)\delta_{s,s^\prime+1} - 2  \nonumber \\
&& - m_f(1-\gamma^5)\delta_{s,0}\delta_{s^\prime,L_s-1} 
       \nonumber \\
&& - m_f(1+\gamma^5)\delta_{s,L_s-1}\delta_{s^\prime,0}
\bigr\}.
\label{eq:DWF}
\end{eqnarray}
The combination of the boundaries at $s=0$ and $s=L_s-1$ and the
negative Wilson mass, $-m_0$, allows unusual massless boundary
states to form, localized on the $s=0$ and $s=L_s-1$ 
four-dimensional boundaries of the five-dimensional problem.  
The result is two types of states, decaying exponentially as one
moves away from 0 or $L_s-1$: massless, right-handed particles
bound to the right-hand wall and massless, left-handed particles
bound to the left-hand wall.  The mixing between these two
boundary states, given by the small overlap of the exponentially
decreasing 5-dimensional wave functions, implies that these
states will actually form a massive four-component fermion.  They
will become literally massless only in the limit of infinite
separation between the walls, $L_s \rightarrow \infty$, a limit
in which this exponentially small overlap vanishes.  

The extra mass term with coefficient $m_f$ in Eq.~\ref{eq:DWF}
explicitly couples the right and left walls and gives these
boundary states an additional mass proportional to $m_f$.  This
extra term provides an explicit mass which is easily adjusted in
contrast to the overlap-induced mass whose dependence on $L_s$ is
not so precisely known.  We will attempt to choose $L_s$ and
$m_f$ so that the effects of the mixing are much smaller than
those of $m_f$, so the mass of the domain-wall states is
explicitly proportional to $m_f$.

\subsection{Chiral Symmetry}

The domain wall fermion formulation described above should become
a theory of massless fermions in the limit $L_s \rightarrow
\infty$.  This is easy to see in the free field case and can be
argued to be true for the interacting theory as 
well \cite{SHAMIR2}.  Heuristically, we observe that if the gauge
coupling is sufficiently small, the relevant gauge configurations
will be smooth on the lattice scale and interact primarily with
the low-energy, domain wall states in exactly the fashion of
four-dimensional QCD.  Of course, the extent to which this is
actually true in a particular calculation for a specific value of
$L_s$, must be studied numerically.

In the usual Wilson formulation of lattice fermions, the axial
part of the underlying $SU(N_f) \times SU(N_f)$ flavor symmetry
is strongly broken by the dimension-five, ``Wilson term'' added
to make the doublers heavy.  A chiral theory is expected to be
found in the low energy portion of the theory provided the 
mass-related hopping parameter $\kappa$ is tuned to its critical
value.  However, in practice, the effects of this Wilson term
have been so large as to obscure the character of the chiral
phase transition and to serious impede the extraction of a
variety of weak matrix elements whose efficient calculation
requires the use of chiral symmetry.

The staggered fermion formulation preserves a single chiral
symmetry, making the chiral phase transition much easier to
simulate.  However, this approach contains lattice artifacts
which break the normal vector flavor symmetries making many
quantities more difficult to interpret than in the Wilson
formulation.

Both the Wilson and staggered formulations of lattice fermions
are expected to show full, physical flavor symmetry only as the
continuum limit is approached.  In contrast, $N_f$ species of
domain wall fermions, as formulated above, support a complete
$SU(N_f) \times SU(N_f)$ flavor symmetry which becomes exact in
the possibly much less demanding $L_s \rightarrow \infty$ limit.

\subsection{Index Theorem on the Lattice}

An important part of the relativistic quantum physics of fermions
is the axial anomaly which is required for proper understanding
of the $\pi^0 \rightarrow \gamma\;\gamma$ decay, the
$\eta^\prime$ mass and the order of the QCD phase transition. 
With both Wilson and staggered fermions, the symmetry generated
by the anomalously non-conserved current is explicitly broken by
order $a$ or order $a^2$ terms in the lattice fermion action. 
Thus, without a careful study of the limit of vanishing lattice
spacing, $a \rightarrow 0$, one is unable to be certain whether a
given anomalous effect results from a lattice artifact or will
survive in the continuum limit.

Happily, this fundamental physical phenomona is also represented
in a new and potentially more reliable fashion by the domain wall
fermion formulation.  As developed by Narayanan and Neuberger,
the domain wall formulation supports a variant of the 
Atiyah-Singer index theorem in the $L_s \rightarrow \infty$
limit, even for finite lattice spacing.  In the continuum, this
theorem relates the winding number of the gauge field background
to the number of zero modes of the Dirac operator and provides
the explicit mechanism for gauge field topology to effect the
physics of quarks.

To understand the connection between gauge field topology and
exact zero modes of the Dirac operator, for domain wall fermions
one begins by viewing the five-dimensional fermion path integral
in a fixed gauge background and for a fixed $L_s$ as representing
both the determinant of $D_{\rm DWF}$ and as a quantum mechanical
expectation value of the transfer matrix $T_5$ for unit
translation in the fifth dimension in the presence of the 
$s$-independent background gauge field:
\begin{equation}
Z(L_s)={\rm det}\{D_{\rm DWF}\}=\langle 0 |T_5^{L_s}|0\rangle.
\label{eq:proj}
\end{equation}
Here, the fermion state $|0\rangle$ is a particular 
four-dimensional fermionic state, determined by the boundary
conditions with half of the available states filled.  In the
limit $L_s \rightarrow \infty$, the right-most factor of
Eq.~\ref{eq:proj} projects onto the eigenstate of $T_5$ with the
largest eigenvalue, $|0^\prime\rangle$.  Thus, we can conclude
that
\begin{equation}
{\rm lim}_{L_s \rightarrow \infty}\;{\rm det}\{D_{\rm DWF}\}
\propto |\langle 0^\prime|0\rangle|^2
\end{equation}
The Dirac operator of our five dimensional formulation will then
have exact zero modes whenever the states $|0\rangle$ and
$|0^\prime\rangle$ have a different number of occupied states
(and are then, necessarily, orthogonal).  This connection of
exact Dirac zero modes with integers defined from the gauge
fields is in close analogy to the continuum Atiyah-Singer
theorem.  In the continuum limit, such zero modes will correspond
to topologically non-trivial gauge configurations.  Furthermore,
these zero modes should continue to exist as lattice corrections
and quantum fluctuation are added \cite{NARAYANAN-VRANAS}.

In the next Section, we will examine the question of how large
$L_s$ must be for these expected near-zero modes to be
recognized.

\section{Fermion Zero Modes}
\label{sec:zero-modes}

In order to investigate the degree to which expected fermion zero
modes will actually be visible for finite $L_s$ and non-zero
lattice spacing, we have looked for zero mode effects in the
background of a discretized, instanton-like gauge
field.\cite{INSTANTON}  We do this by evaluating the 
volume-averaged, chiral condensate as a function of the explicit
quark mass, $m_f$.  In the continuum, the chiral condensate is
easily related to the spectrum of Dirac eigenvalues through the
Banks-Casher formula:
\begin{equation}
\langle\bar\psi\psi\rangle= 
                  -m\int_{-\infty}^\infty{\rho(\lambda,m)d\lambda
                                         \over \lambda^2 + m^2}
\label{eq:banks}
\end{equation}
where $\rho(\lambda)$ is the density of Dirac eigenvalues
corresponding to the gauge configuration or ensemble of
configurations used to compute $\bar\psi\psi$ and $m$ is the bare
quark mass.

\begin{figure}[t]
\vskip -0.75in
\epsfxsize=3.5in
\epsfbox{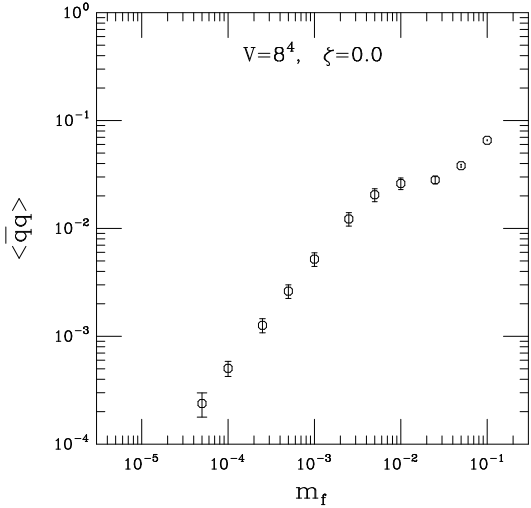}
\vskip -0.25in
\caption{The chiral condensate computed on an instanton-like
background evaluated using staggered fermions.  The small 
inflection interrupting the steady decrease of
$\langle\bar\psi\psi\rangle$ with decreasing $m_f$ is caused by
the expected ``zero-mode'', shifted far from zero by this coarse
$8^4$ lattice.}
\label{fig:pbp_stg}
\vskip 0.0in
\end{figure}

Eq.~\ref{eq:banks} implies that the presence of a zero mode,
$\rho(\lambda)= Z\delta(\lambda) + ...$ implies a $Z/m$
divergence in $\bar\psi\psi$ as $m$ approaches zero.  Such 
zero-mode effects can be easily studied 
\cite{ZMODES,LAT98-TOPOLOGY} by computing $\bar\psi\psi$ for a
fixed gauge configuration, constructed to be close to a Belavin,
Polyakov, Schwartz, Tyupkin instanton.  Such a configuration will
have an exact zero mode for the continuum Dirac operator.  In
Figure~\ref{fig:pbp_stg}, we show $\bar\psi\psi$ as a function of
quark mass computed in this fixed, instanton-like background
using the staggered Dirac operator.  Instead of the desired
$1/m_f$ divergence as $m_f\rightarrow 0$, one sees a weak
inflection around $m_f \approx 10^{-2}$ suggesting that the
lattice spacing has shifted the zero mode from zero to
approximately this value.  Even this weak signal essentially
disappears if even 10\% noise is superimposed on the original
instanton-like background.

\begin{figure}
\vskip -0.75in
\epsfxsize=3.5in
\epsfbox{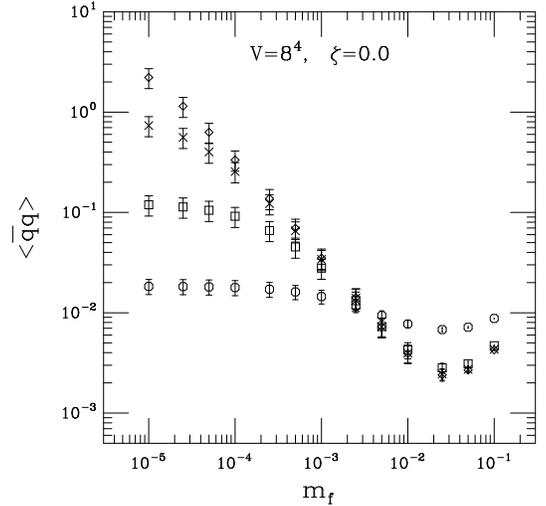}
\vskip -0.25in
\caption{The chiral condensate computed on the instanton-like
background of Figure~1, now evaluated using domain wall fermions. 
Results for four different choices of $L_s$ are shown: $L_s=4$
(circles), 6 (squares), 8 (crosses) and 10 (diamonds).  Now the
expected, zero-mode induced $1/m_f$ behavior can be easily seen
for $L_s>4$ and $m_f \le 10^{-2}$.}
\label{fig:pbp_dwf}
\vskip 0.0in
\end{figure}

The behavior for the domain wall Dirac operator in this same
background, shown in Figure~\ref{fig:pbp_dwf}, is dramatically
different.  There one sees the expected decrease in
$\bar\psi\psi$ as $m_f$ decreases, until $m_f \approx 2\; 10^{-
2}$, at which point a clear $1/m_f$ signal is seen, even for the
curve corresponding to $L_s$ as small as 6.  For finite $L_s$
this $1/m_f$ behavior is cut off and $\bar\psi\psi$ becomes
constant as $m_f$ falls below the residual mass introduced by
direct mixing between the two overlapping domain wall states.  As
can be seen in the figure, this effect moves to much smaller
$m_f$ as $L_s$ increases.  Equally important, the appealing
picture shown in Figure~\ref{fig:pbp_dwf} changes very little as
noise is superimposed on the instanton-like, background solution. 

Thus, it appears that the domain wall formulation will allow us
to simulate fermions at fixed lattice spacing and not-too-large
$L_s$ which show both the full $SU(2)\times SU(2)$ flavor/chiral
symmetry and also the important connection between lattice
topology and fermion zero-modes which underlies the physics of
the axial anomaly.

\section{QCD at Zero Temperature}
\label{sec:zero-temp}

\subsection{Hadronic Spectrum}

\begin{figure}
\epsfxsize=3.0in
\epsfbox{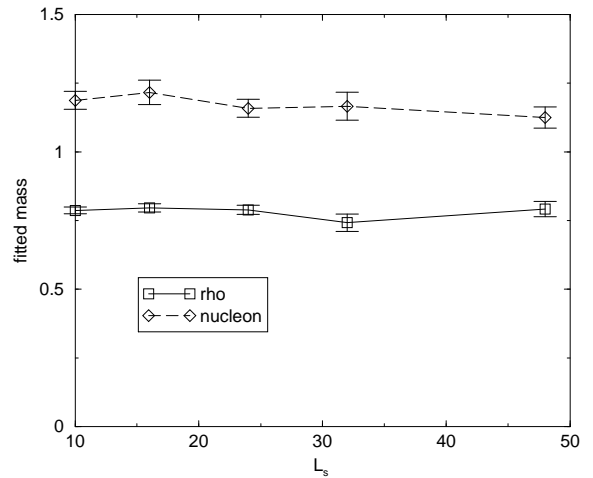}
\caption{The $\rho$ and nucleon masses, extrapolated to $m_f=0$,
plotted as a function of $L_s$, the lattice extent in the fifth
dimension. These masses were computed with domain wall height
$m_0=1.65$ from quenched, $\beta=5.85$, $8^3\times 32$
configurations.}
\label{fig:nucl_rho}
\end{figure}

As our first test of these ideas we have carried out a series of
careful, quenched calculations on small $8^3\times 32$ lattices
for a variety of values of $L_s$ and $m_f$.\cite{LAT98-QUENCHED}
Using physical quark operators whose left- and right-handed
components were obtained by evaluating the 5-dimensional fermion
fields on the $s=0$ and $s=L_s-1$ hyperplanes, we had little
difficulty computing hadron masses following the usual methods. 
In Figure~\ref{fig:nucl_rho}, we show the resulting $\rho$ and
nucleon masses, extrapolated to $m_f=0$, as a function of $L_S$. 
Although we have examined quite large domain wall separations, up
to $L_s=48$, it is clear from the figure, that we could have
extracted accurate results from the smaller $L_s=10$ or 16
calculations.

\begin{figure}
\epsfxsize=3.0in
\epsfbox{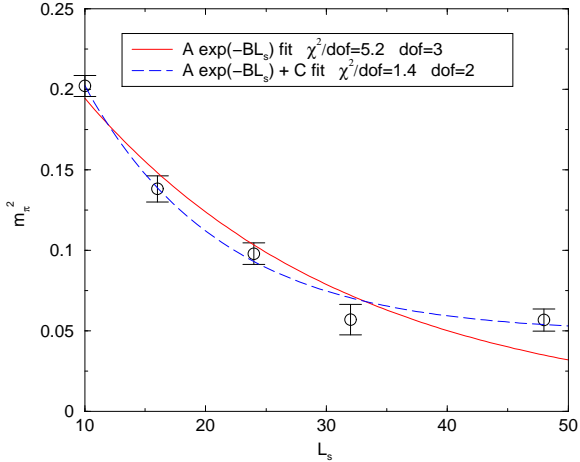}
\caption{The $\pi$ mass, extrapolated to zero quark mass, plotted
as a function of $L_s$, again for quenched, $\beta=5.85$,
$8^3\times 32$ configurations.  The failure of $m_\pi$ to
approach zero as $L_s \rightarrow \infty$, as required by the
Goldstone theorem, is most likely a simple, finite-volume
effect.}
\label{fig:pi_mass}
\end{figure}

The corresponding $L_s$ behavior of the pion mass is shown in
Figure~\ref{fig:pi_mass}.  In contrast to the case of the nucleon
and the $\rho$, we see significant dependence on $L_s$, even at
the largest values of $L_s$.  This contrast can be easily
explained if we hypothesize that the largest effect of finite
$L_s$ is to give an additional $L_s$-dependent contribution to
the quark mass.  Such an effect will be much more visible for the
very light pion than for the heavier $\rho$ or nucleon.  Using
the observed $m_f$ sensitivity (not shown here) of $m_\rho$ and
$m_\pi$, one can conclude that the 100\% change in
$m_\pi^2|_{m_f=0}$ seen as $L_s$ increases from 16 to 48, would
amount to only a 4\% effect on $m_\rho|_{m_f=0}$, an effect
hidden by our errors.  Of greater potential interest is the
failure of $m_\pi$ to vanish as $L_s \rightarrow \infty$.  While
this may be a new feature of the quenched approximation made
visible by the domain wall formalism, it is very likely a simple,
finite-volume effect.  This non-vanishing contribution to $m_\pi$
is an appreciable fraction of $m_\pi$ only for very small quark
masses, $m_f \approx 0.02$.  However, for such light quarks the
product $m_f \; \langle\bar\psi\psi\rangle\; V = 0.02 \cdot
0.0019 \cdot 8^3 \times 32 = 0.7$, a clear warning that the
Goldstone phenomena will begin to be influenced by our finite
$8^3\times 32$, 
space-time volume.  

\subsection{Chiral Condensate}

Lacking the fermion determinant, the quenched approximation can
potentially yield gauge configurations with small or vanishing
Dirac eigenvalues.  As was discussed in 
Section~\ref{sec:zero-modes}, such configurations could lead to 
an unphysical, $1/m$ divergence in the chiral condensate. 
However, such behavior could be easily suppressed by lattice
artifacts for staggered fermions and may be reduced when
``exceptional'' configurations are discarded for Wilson fermions.

\begin{figure}
\epsfxsize=3.0in
\epsfbox{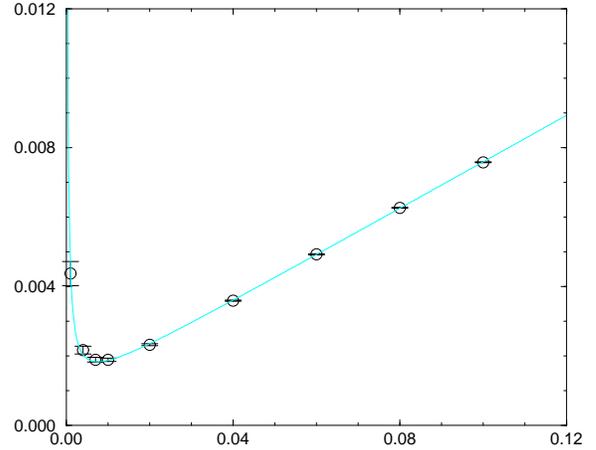}
\caption{The chiral condensate is plotted as a function of quark
mass from the same quenched, $\beta=5.85$ $8^3\times 32$
calculation.  The clear $1/m_f$ behavior is a new, unphysical
feature of the quenched approximation, resulting from the absence
of the fermion determinant.}
\label{fig:pbp_nt32}
\end{figure}

We have explicitly searched for such effects in a
$\beta=5.85$, quenched simulation and found them quite
easily.\cite{LAT98-PBP}  Our results for a  $8^3\times 32$
lattice are shown in Figure~\ref{fig:pbp_nt32}.  One sees a
clearly $1/m_f$ divergence for small $m_f$.  The $1/m_f$
coefficient is $3.8(3)\;10^{-6}$.  If this effect comes only from
exact zero modes, whose number increases as $V^{1/2}$, we should
expect this behavior to be less pronounced on larger volumes.  In
fact, a similar, $16^3\times32$ calculation also shows this
$1/m_f$ behavior, but with a much smaller coefficient,
$0.6(1)\;10^{-6}$.  Thus, while one should worry that
conventional quenched chiral extrapolations do not allow for such
behavior, this new effect may not be important for quenched
calculations performed on large physical volumes.

\section{QCD near $T_c$}

Because of the importance of chiral symmetry for the QCD phase
transition, one of the most interesting applications of domain
wall fermions may be to studies of QCD thermodynamics.

\subsection{The quenched Chiral Condensate}

\begin{figure}
\epsfxsize=3.0in
\epsfbox{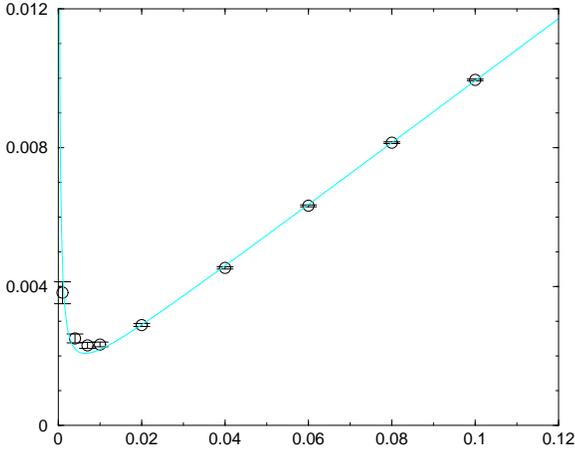}
\caption{The chiral condensate is plotted as a function of quark
mass, computed using pure gauge configurations on a $16^3\times
4$ lattice, just above the deconfining phase transition for
$\beta=5.71$.  There is an new $1/m_f$ divergence visible for
small mass as well as a non-zero constant piece in
$\bar\psi\psi$.  These appear to be new, volume-independent
terms, first visible when domain wall fermions are used.}
\label{fig:pbp_nt4}
\end{figure}

We will first continue the discussion above of possible quenched,
$1/m_f$ behavior for $T \ge T_c$.\cite{LAT98-PBP}  Here, one
expects to see such singular behavior and a $1/m_f$ term that may
be non-zero, even in the limit of infinite volume.\cite{OMEGA} 
While earlier staggered fermion calculations have not seen this
effect,\cite{OMEGA,ADRIAN} it is very visible in the $\beta=5.71$
results shown in Figure~\ref{fig:pbp_nt4}.  A companion
calculation on a larger $32^3\times 4$ lattice shows $1/m_f$
behavior with the same coefficient.  Thus, by using a fermion
formulation more sensitive to the effects of zero modes, we have
discovered that the quenched chiral condensate shows quite
striking behavior.  Rather than vanishing for small $m_f$ in the
deconfined region, as would be expected if quenched chiral
symmetry were restored above $\beta_c$, the chiral condensate
diverges as $1/m_f$.  In addition, as can be seen in
Figure~\ref{fig:pbp_nt4}, there is a non-zero, $m_f=0$ intercept
even if the $1/m_f$ term is removed.  Clearly the old picture of
quenched chiral symmetry restoration above $\beta_c$ is seriously
incomplete.

\subsection{$SU(2)\times SU(2)$ Chiral Transition}

\begin{figure}
\vskip -0.5in
\epsfxsize=3.5in
\epsfbox{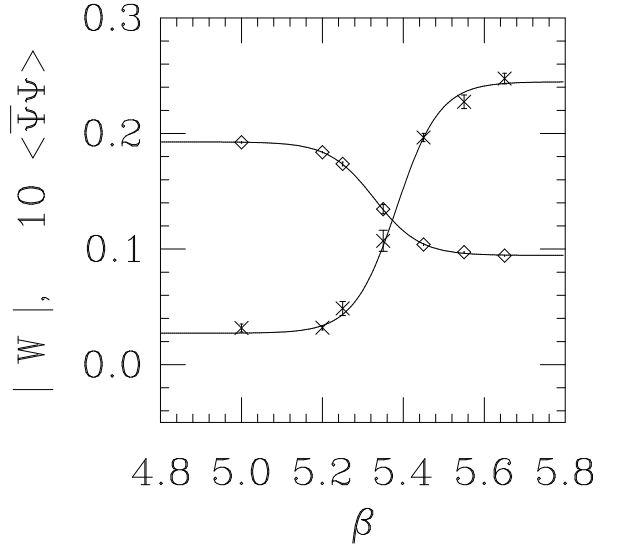}
\caption{The chiral condensate and Wilson line are plotted as a
function of $\beta$.  These values were obtained in a full QCD,
$N_f=2$, $m_f=0.1$ calculation on an $8^3\times 4$ lattice. 
Behavior characteristic of a phase transition with $\beta_c
\approx 5.35$ is seen.}
\label{fig:transition}
\end{figure}

Now let us examine the physical, 2-flavor QCD phase transition
using domain wall fermions.\cite{LAT98-THERMO}  We have done a
series of calculations on $8^3\times 4$ lattices with a variety
of values of $\beta$, $L_s$ and domain wall heights.  In
Figure~\ref{fig:transition}, we show the Wilson line expectation
value and the chiral condensate computed with a domain wall
height $m_0=1.90$, $L_s=12$ and quark mass $m_f=0.1$.  One sees
the expected cross-over behavior as $\beta$ increases, suggesting
that a full QCD, thermodynamic calculation using domain wall
fermions is may well be possible.  We have seen very similar
behavior for $m_0$ varying between 1.65 and 2.15, suggesting that
we are seeing stable, 2-flavor behavior for $m_0=1.9$, although
more study of this question is warranted.

\begin{figure}
\vskip -0.5in
\epsfxsize=3.5in
\epsfbox{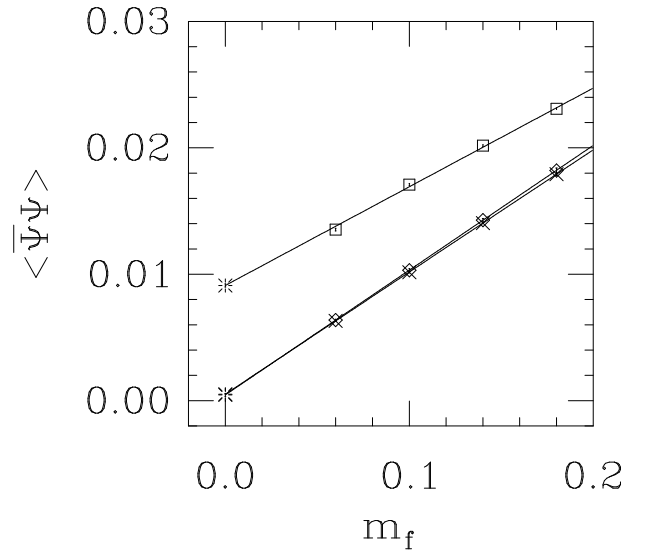}
\caption{The chiral limit of the $\langle\bar\psi\psi\rangle$ in
full, 2-flavor QCD for $L_s=16$ and $m_0=1.9$  Shown are points
below the transition, $\beta=5.20$, $8^3\times4$(squares) and
points above the transition, $\beta=5.45$, $8^3\times4$(diamonds)
and $16^3\times4$(crosses).}
\label{fig:pbp_pmtc}
\end{figure}

Of special interest is the degree of chiral symmetry present on
either side of this transition region.  This is examined in
Figure~\ref{fig:pbp_pmtc} where the limit $m_f \rightarrow 0$ is
displayed using a series of full QCD calculation with quark
masses $m_f=0.06$, 0.10, 0.14 and 0.18.  The domain wall
formulation appears to have allowed a calculation, for the first
time, in which a transition is seen between a phase with
spontaneously broken chiral symmetry ($\beta=5.20$) and one with
what should be full, restored, $SU(2)\times SU(2)$ symmetry
($\beta=5.45$).  Calculations are now underway to study this
system further, examining the properties of the QCD phase
transition in greater detail for larger spatial volumes.

\subsection{Anomalous Symmetry Breaking}

\begin{figure}
\epsfxsize=3.0in
\epsfbox{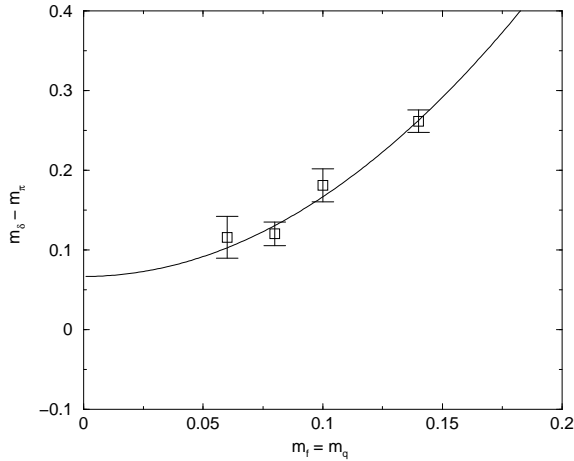}
\caption{The anomalous difference of the screening masses for the
$\pi$ and $\delta$ are plotted as a function of quark mass. 
These calculations were performed with two flavors of dynamical
quarks on a $16^3\times 4$ lattice, at $\beta=5.40$, just of
above the transition, using $L_s=16$ and $m_0=1.9$}
\label{fig:anomaly}
\end{figure}

As a final topic, we attempt to exploit the sensitivity of the
domain wall fermion formulation to topological effects, and
examine the degree of anomalous symmetry breaking slightly above
the phase transition.\cite{LAT98-THERMO}  The current results of
this study are shown in Figure~\ref{fig:anomaly}.  We plot the
difference of two screening masses, computed at $\beta=5.40$,
just above the transition region, evaluated on a $16^3\times 4$
lattice, using $L_s=16$ and $m_0=1.9$.  Here we are subtracting
the screening mass $m_\delta$, extracted from the exponential
damping of correlation functions computed from the operator
$(\bar\psi\vec\tau\psi)(x)$, and $m_\pi$ computing using the
operator $(i\bar\psi\vec\tau\gamma^5\psi)(x)$.  The $2\times 2$
matrices, $\vec\tau$, are the usual Pauli, isospin generators.

Because the $\delta$ and $\pi$ correlation functions are related
by the anomalous chiral symmetry, any difference between these
two screening masses in the limit $m_f\rightarrow 0$, is an
unambiguous measure of anomalous chiral symmetry breaking.  Since
these screening masses are both about 1 in lattice units,
Figure~\ref{fig:anomaly} shows a small ($\approx 5\%$), but
possibly significant effect.  The present data, shown in the
figure, give an intercept of 0.067(1.8)when fitted to the
expected quadratic $m_f$ dependence.  While we cannot be certain
we have seen the sought-after evidence of anomalous symmetry
breaking (the sort required to predict a second-order QCD phase
transition), we can assert with confidence that the effect is
quite small.

\section{Conclusion}

We have demonstrated that the domain wall formulation of lattice
fermions can be used to study both zero temperature, quenched
hadron spectroscopy and the QCD phase transition for both zero
and two flavors.  This new method appears to realize accurate
chiral symmetry at fixed, non-zero lattice spacing, and does not
require unreasonably large domain wall separation in the fifth
dimension.  This method offers a very promising new approach to
the study of chiral symmetry using lattice methods.

\section*{Acknowledgements}
This work was supported in part by the U.S. Department of Energy. 
One of us (G.S.) acknowledges the support of the Max Kade
Foundation.  We have benefited from discussions with a number of
people included T.~Blum, R.~Edwards, R.~Narayanan, H.~Neuberger,
and Y.~Shamir.

\end{document}